\documentclass[twocolumn,10pt,pra,aps]{revtex4-1}
\def\by#1{#1,}
\def\and{and }
\def\yr#1{{(#1)}}
\def\paper#1{#1}
\def\jour#1{{\it #1}}

\def\vol#1{{\bf #1},}
\def\issue#1{}
\def\pages#1{\hbox{#1},}

\usepackage{amsmath}
\usepackage{amsfonts}
\usepackage{amssymb}
\usepackage{graphicx}
\usepackage{appendix}
\usepackage{placeins}

\newcommand{\pdiff}[3][]{\dfrac{\partial^{#1} #2}{\partial {#3}^{#1}}}
\begin{document}

\title{Breakup of oil slicks on wavy water surfaces}

\author{Alex V. Lukyanov}
\email{corresponding author,\\ a.lukyanov@reading.ac.uk}
\affiliation{School of Mathematical and Physical Sciences, University of Reading, Reading, RG6 6AX, UK}

\author{Hanan Hozan}
\affiliation{School of Mathematical and Physical Sciences, University of Reading, Reading, RG6 6AX, UK}

\author{Georgios Sialounas}
\affiliation{School of Mathematical and Physical Sciences, University of Reading, Reading, RG6 6AX, UK}

\author{Tristan Pryer}
\affiliation{Department of Mathematical Sciences, University of Bath, Bath, BA2 7AY, UK}

\begin{abstract}
We hypothesize that the spread of oil slicks on the water's surface during oil spills is significantly influenced by water wave motion at the initial or intermediate spreading stages, well before emulsification processes have a substantial impact on the oil film's state. We demonstrate that the spreading dynamics of an oil slick on the water surface are facilitated by water waves, employing the thin film approximation. It is shown that water wave motion can rapidly deplete any oil slick, reducing the oil layer's thickness to nearly zero. This mechanism may act as a precursor to emulsification processes, leading to the accelerated depletion of oil spills into a distribution of droplets that form an emulsion.
\end{abstract}

\maketitle
 
\section{Introduction}

The dynamics of oil spills on the sea surface represent a significant phenomenon with practical implications~\cite{GasIndustry2016}. Initial theoretical studies, supplemented by laboratory experiments, concentrated on the simplified scenarios of oil spreading. These studies primarily examined the dynamics of liquid films on nearly flat water-oil interfaces—essentially on the surface of calm seas—to delineate the fundamental physical mechanisms involved and to identify the various stages of oil spreading~\cite{Shewmaker1954, Fay1969, Fay1971, Hoult1972, Cox1980, Phillips1996}.

In laboratory settings, it has been determined that spreading can be classified into three consecutive, distinct regimes based on the dominant forces involved. Specifically, these include a brief, initial gravity-inertial phase (lasting a few minutes), followed by a gravity-viscous spreading regime (lasting several hours), and culminating in the viscous-surface tension regime~\cite{Fay1971, Hoult1972}.

Oil spreading under realistic environmental conditions is a significantly more complex phenomenon, influenced by a multitude of natural factors often absent in laboratory experiments. These factors include water wave motion, winds, currents, general turbulence, oil evaporation, dispersion, subsequent emulsification due to turbulent motion, biodegradation, and interactions between oil and the shoreline.

Particularly important is the role of surface tension (capillary effects), which predominates in the final spreading stage under laboratory conditions. This factor is crucial as it directly governs the emulsification process at the microscale.

Therefore, in efforts to forecast the overall environmental impact, factors such as rough sea conditions have historically been incorporated empirically into models. This approach emphasizes simplified and empirical methodologies, yet ensures the inclusion of all pertinent natural factors at realistic scales~\cite{Murray1972, Sahota1979, Raj1979, Marsh1981, Huang1983, Lehr1984, Lehr1984PB, Areview1996}.

Particularly, recent research efforts have been concentrated on understanding the emulsification processes and the large-scale behavior of water-oil emulsions formed as a result. This includes focusing on the dynamics and monitoring of the distribution of oil micro-droplets on a large scale~\cite{Markatos2005, Li2011, Skartlien2013, Soloviev2016, Geng2016, Nissanka2017, Nissanka2018, Fingas2018, Garcia2020, Zhao2020, Sundhar2023}.

\begin{figure}[ht!]
\begin{center}
\includegraphics[trim=0.3cm 2.cm 1cm -0.5cm,width=\columnwidth]{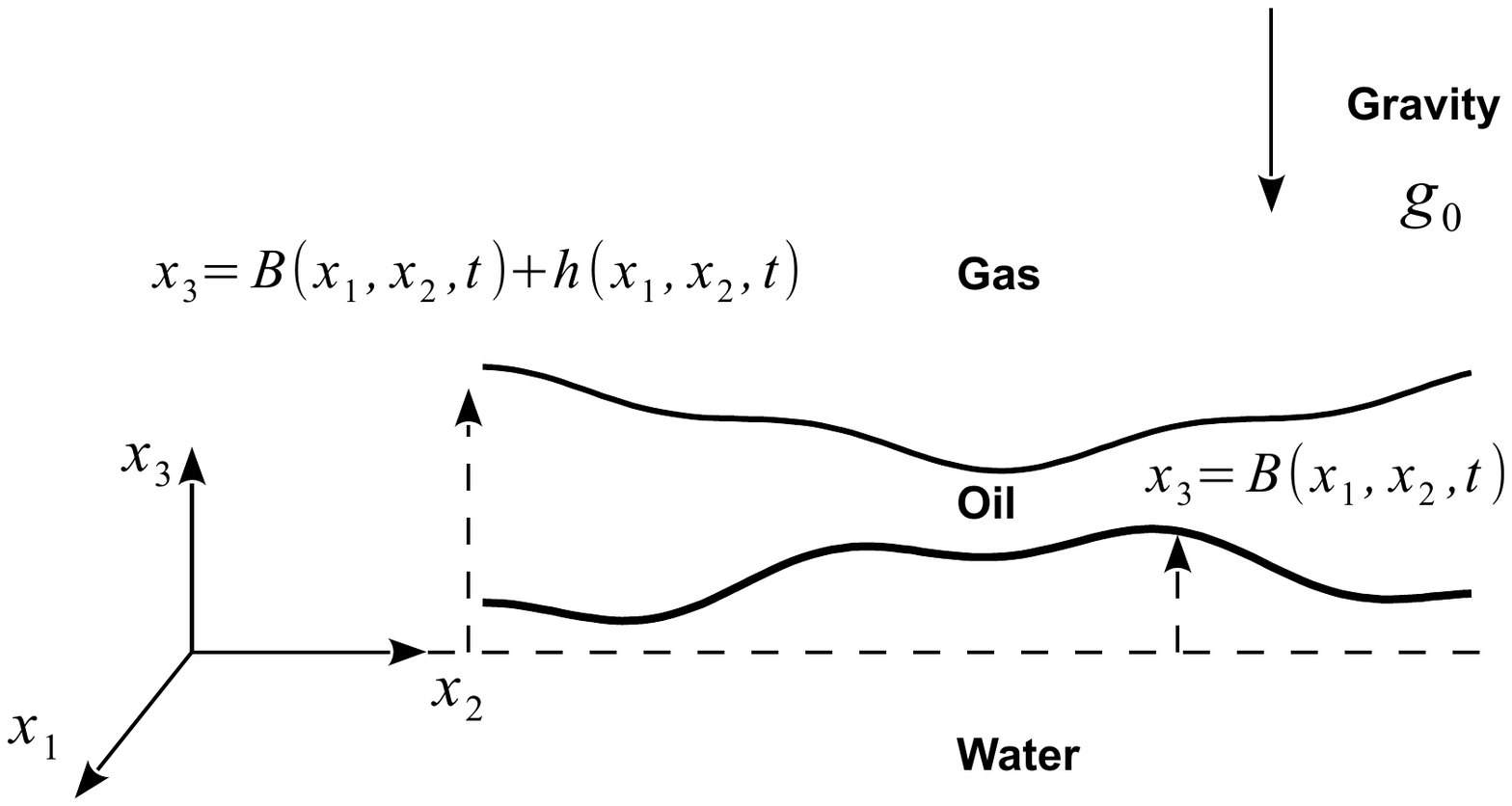}
\end{center}
\caption{Illustration of the oil layer on the water surface problem geometry.} 
\label{Fig1}
\end{figure}

 In this study, we aim to provide a more rigorous and quantitative analysis of the influence of water-wave motion on the spreading of oil slicks. Specifically, we explore the potential for the breakup of oil slicks by surface water waves at the initial stages of oil-film spreading, essentially at the onset of the gravity-viscous regime.

It is well-established that factors such as winds, currents, and water turbulence significantly impact the spreading rate~\cite{Zhao2020}. Notably, the spreading rates predicted by the original Fay-Hoult theory, which assumes calm sea conditions, are markedly lower than those typically observed in real-world conditions~\cite{Zhao2020}.

In this paper, we demonstrate that under relatively modest sea surface conditions, a continuous oil slick can rapidly disintegrate into surface fractions significantly before the emulsification process begins to exert a noticeable effect, namely at the very start of the gravity-viscous regime.

The depletion process, it appears, could significantly influence the further evolution of the oil slick. The formation of void areas is expected to considerably enhance emulsification processes along substantially longer boundaries. Furthermore, we will show that the spreading rate is impacted even in the absence of depletion.

In the subsequent sections, we will introduce a mathematical model based on the thin film approximation. This model will be analyzed numerically to investigate various dynamics of the oil slick.

\section{Problem formulation and the mathematical model}

Consider an oil layer with density $\rho$ and zero shear rate dynamic viscosity $\mu$, spreading on the water's wavy surface, as depicted in Fig. \ref{Fig1}. In this study, we assume the oil is Newtonian, although it's noted that various crude oil compositions could exhibit non-Newtonian behavior.

The positions of the oil-gas and oil-water interfaces are denoted by $x_3=B(x_1,x_2,t)$ and $x_3=B(x_1,x_2,t) + h(x_1,x_2,t)$, respectively, where $h$ represents the thickness of the oil layer.

The mathematical framework is succinctly introduced as the thin-film lubrication approximation~\cite{Matar2009}. We will focus on details pertinent to the specific spreading regime under consideration. In the thin-film approximation, the model distinguishes between two length scales: $L$ in the $x_{1,2}$-directions, which might represent the diameter of the oil slick or the wavelength of surface disturbances, and $H$ in the $x_3$-direction, corresponding to the oil layer's thickness.

The ratio of these scales, defined as $\frac{H}{L} = \varepsilon$, is considered a small parameter, $\varepsilon \ll 1$. This assumption is based on the typical ranges of oil-slick characteristic lengths, with $1,\text{m} \leq L \leq 100,\text{m}$ covering either the oil slick's size or the scale of perturbations, and $10^{-4},\text{m} \leq H \leq 10^{-2},\text{m}$ representing the thickness of the oil layer.

In our analysis, we employ several simplifying assumptions for clarity and focus. Firstly, we assume the presence of long water waves, where the characteristic wavelength $L$ significantly exceeds the thickness of the oil layer. In this context, surface tension effects can be disregarded, as $L$ vastly surpasses the capillary length $L \gg L_c=\sqrt{\frac{\gamma}{\rho g_0}}$, which is approximately 2 mm for most liquids. Here, $\gamma$ represents the surface tension coefficient, and $g_0$ denotes the acceleration due to gravity.

Our second assumption is inherently linked to the first. Within the long-wavelength regime, we presume that the oil layer exerts no retroactive effect on water movement. This assumption of negligible feedback is particularly reasonable when considering the spreading process over a non-stationary water interface.

The governing equations in the thin film approximation are obtained by using the small parameter $\varepsilon$ and averaging the Navier-Stokes equations over the oil layer, that is by introducing the liquid fluxes 
$$
q_{1,2} = \int_B^{h+B}\, v_{1,2}\, dx_3,
$$
where $v_{1,2}$ are the components of the velocity vector.

The original system of the Navier-Stokes equations 
\begin{align}
\rho\left(\frac{\partial v_i}{\partial t}+v_{\ell}\frac{\partial v_i}{\partial x_{\ell}}\right)&=-\frac{\partial p}{\partial x_i}+\mu \frac{\partial^2 v_i}{\partial x_j^2}+F_i,\label{momentumConservation}\\
\frac{\partial v_k}{\partial x_{k}}&=0,\label{massConservation}
\end{align}
where $p$ is pressure and ${\bf F}=\{0,0,\rho g_0\}$ are the gravity body-force components, is brought into a non-dimensional form by introduction of the reduced variables
\begin{equation}
\widetilde{x}_{1,2}=x_{1,2}/L,\quad\widetilde{x}_3=x_3/H=x_3/\varepsilon L,
\end{equation}
\begin{equation}
\widetilde{v}_{1,2}=v_{1,2}/U,\quad\widetilde{v}_3= v_3/\varepsilon U,\quad\widetilde{t}=t/t_0,
\end{equation} 
where $U$ is the characteristic velocity and $t_0=L/U$. The pressure in the system is assumed to be dominated by viscous contributions, so that it was normalised by $p_0=\frac{\mu U}{\varepsilon H}$, $\tilde{p}=p/p_0$. 

The boundary conditions to (\ref{momentumConservation}) - (\ref{massConservation}) on the oil-gas interface $x_3=B(x_1,x_2,t) + h(x_1,x_2,t)$ are negligible stress tangential to the surface area
\begin{equation}
\label{FreeS-1}
n_j\, (\delta_{il}-n_{i} n_{l}) \left (-p\delta_{ij}+\left(\frac{\partial v_i}{\partial x_j} + \frac{\partial v_j}{\partial x_i}\right)\right)=0.
\end{equation}
and the continuity of the normal stress component
\begin{equation}
\label{FreeS-2}
n_i n_j \left((p_e-p)\delta_{ij } + \left(\frac{\partial v_i}{\partial x_j} + \frac{\partial v_j}{\partial x_i}\right) \right) = \frac{\varepsilon^2}{Ca} \frac{\partial n_k}{\partial x_k}.
\end{equation}
Here, $p_e$ is the external gas pressure, $Ca=\frac{\mu U}{\gamma}$ is the capillary number, $n_j$ are the components of the normal vector to the interface pointing into the liquid and $\delta_{ij}$ is the Kronecker-Delta. 

In the thin film approximation, neglecting surface tension effects, conditions (\ref{FreeS-1}) and (\ref{FreeS-2}) are reduced to
\begin{equation}
\label{FreeS-1SF}
 \frac{\partial v_1}{\partial x_3}=0, \quad  \frac{\partial v_2}{\partial x_3}=0
\end{equation}
and
\begin{equation}
\label{FreeS-2SF}
p=p_e
\end{equation}
respectively.

On the oil-water interface $x_3=B(x_1,x_2,t)$, the no-slip condition is assumed in the tangential to the interface direction, such that the oil motion is driven by the water-wave velocity $V_i$,
\begin{equation}
\label{noslip}
 (\delta_{il}-n_{i} n_{l}) v_i = (\delta_{il}-n_{i} n_{l}) V_{i}(x_1,x_2,t).
\end{equation}
In the thin film approximation, the last condition is equivalent to
\begin{equation}
\label{noslipr}
v_i =  V_{i}(x_1,x_2,t)\quad i=1,2.
\end{equation}

The dynamics of both interfaces is controlled by kinematic boundary conditions, that is

\begin{equation}
\label{kctfint}
\left. v_3\right|_{x_3=B}=\pdiff{B}{t} + v_1\pdiff{B}{x_1} + v_2\pdiff{B}{x_2}.
\end{equation}
and
\begin{equation}
\label{kctf}
\left. v_3\right|_{x_3=h+B}=\\
\end{equation}
$$
\pdiff{(h+B)}{t} + v_1\pdiff{(h+B)}{x_1} + v_2\pdiff{(h+B)}{x_2}.
$$

After the averaging over the oil layer by using the Karman-Pohlhausen ansatz, 
\begin{equation}
\label{ansatz}
v_{1,2} = V_{1,2} - 
\end{equation}
$$
\frac{3}{2}\frac{q_{1,2}-V_{1,2}h}{h^3}\left\{ x_3^2 - 2(h+B)x_3 +B(B+2h) \right\},
$$
which satisfies the boundary conditions,
one gets
\begin{equation}
\label{LE2}
\frac{\partial h}{\partial t}  + \frac{\partial q_1}{\partial x_1} + \frac{\partial q_2}{\partial x_2} = 0,
\end{equation}
$$
\frac{\partial q_1}{\partial t} +  \frac{\partial }{\partial x_1} \left(\frac{6}{5} \frac{q_1^2}{h} -\frac{2}{5}(q_1 V_1) + \frac{1}{5}V_1^2 h \right) +
$$
\begin{equation}
\label{GE2D3s}
 \frac{\partial }{\partial x_2} \left( \frac{6}{5} \frac{q_1q_2}{h} -\frac{1}{5}(q_1 V_2 + q_2 V_1) + \frac{1}{5} V_1 V_2 h \right) = 
\end{equation}
$$
- Ka\, h\, \pdiff{(h+B)}{x_1} -\frac{3}{Re}\frac{q_{1}-V_1 h}{h^2}
$$
and
$$
 \frac{\partial q_2}{\partial t} + \frac{\partial }{\partial x_1} \left( \frac{6}{5} \frac{q_1q_2}{h} -\frac{1}{5}(q_1 V_2 + q_2 V_1) + \frac{1}{5} V_1 V_2 h + \right) +
$$
\begin{equation}
\label{GE2D4s}
 \frac{\partial }{\partial x_2} \left( \frac{6}{5} \frac{q_2^2}{h} -\frac{2}{5}(q_2 V_2) + \frac{1}{5} V_2^2 h\right)=
\end{equation}

$$
- Ka\, h\,  \pdiff{(h+B)}{x_2}  -\frac{3}{Re}\frac{q_{2}-V_2 h}{h^2}.
$$
Here, $Re=\varepsilon \frac{\rho U H}{\mu}$ and $Ka=\frac{g_0 H}{U^2}$.

In the limit of small Reynolds number $Re\ll 1$, the system of equations is reduced to a single non-linear advection-diffusion equation

\begin{equation}
\label{LE2TH}
 \frac{\partial h}{\partial t}  +  \frac{\partial q_1}{\partial x_1} + \frac{\partial q_2}{\partial x_2}  = 0,
\end{equation}
$$
q_i = -  \frac{\alpha_g}{3}\pdiff{(h+B)}{x_i} h^3 +V_i h,
$$
where $\alpha_g = \frac{\rho g_0 \varepsilon H^2}{\mu U}$.

As the oil slick is finite in the horizontal dimensions, we assume that there is a smooth boundary $\Gamma$ of the domain, where
$$
\left. h\right|_{\Gamma} = 0.
$$

The motion of the boundary in the perpendicular to the boundary direction is described by means of the flux at the boundary. That is the boundary velocity $\bf u$
$$
{\bf u}\cdot{\bf n}_{\Gamma} = {\bf q} \cdot{\bf n}_{\Gamma}\, h^{-1},
$$
where ${\bf n}_{\Gamma}$ is the normal vector to $\Gamma$ in the $(x_1,x_2)$-plane.

\section{Low Reynolds number regime of spreading}

Consider the spreading regime at $Re\ll 1$. To analyse the effects of the external forcing due to the wave motion, we consider a one-dimensional problem posed on a compact manifold $x_l\le x \le x_r$
\begin{equation}
\label{LE2TH1D}
 \frac{\partial h}{\partial t}  +  \frac{\partial q}{\partial x}  = 0
\end{equation}
$$
q = -  \frac{\alpha_g}{3}\pdiff{(h+B)}{x} h^3 +V h
$$
with the boundary conditions
$$
h_{x_l} = h_{x_r} = 0.
$$
The domain boundaries are moving with the rate
\begin{equation}
\label{BVL}
\frac{d x_l}{dt} = \left. \frac{q}{h}\right|_{x=x_l} = -  \frac{\alpha_g}{3}\pdiff{(h+B)}{x} h^2 + V  
\end{equation}
and
\begin{equation}
\label{BVR}
\frac{d x_r}{dt} =\left. \frac{q}{h}\right|_{x=x_r} = -  \frac{\alpha_g}{3}\pdiff{(h+B)}{x} h^2 +V.
\end{equation}

The water-oil interface perturbations, that is functions $B$ and $V$, are taken as harmonics of the deep-water waves in the linear approximation 
\begin{equation}
\label{DPWTH}
B(x,t) = \sum_j B_0^j\sin(k_j x - \omega_j t), 
\end{equation}
$$
V(x,t) =\sum_j  B_0^j \varepsilon \omega_j \sin(k_j x - \omega_j t).
$$ 

Here $B_0^j$ is the amplitude, $\omega_j$ is the frequency and $k_j$ is the wave vector. In non-dimensional form, the dispersion relation is
\begin{equation}
\label{DispR}
\omega_j^2 = \varepsilon^{-1}\, Ka\, k_j.
\end{equation}

In the absence of perturbations, the governing equation is reduced to a non-linear diffusion equation, which has self-similar solutions (subject to $h=0$ at the moving boundary) of the form
\begin{equation}
\label{SSS}
h(x,t) = \frac{A_s}{ t^{\frac{1}{n(3n+2)}} }\left(C-\left(\frac{x}{t^{\frac{1}{3n+2}}}\right)^2\right)^{1/3},
\end{equation}
representing a drop-like shape of the oil slick spreading out and flattening over time. Here, $A_s = \left( \frac{3}{2(3n+2)} \right)^{1/3}$ and $C$ is the parameter, which is defined by the total initial amount of the liquid. Parameter $n$ is an integer number defined by the problem dimension, which is $n=1$ for one-dimensional spreading and $n=2$ in the case of two-dimensional axisymmetric spreading. The self-similar solutions are attractors, so that an arbitrary initial profile may ultimately, and quite quickly, evolve to one of them. 

If we consider the axisymmetric 2D case, the algebraic form (\ref{SSS}) suggests that the boundary (radius $r$) of the domain is moving according to 
$$
r\propto t^{\alpha}, \quad \alpha = 1/(3n+2) = 1/8.
$$
which is well in the range of the exponents found in droplet spreading experiments~\cite{Marmur1981} over solid surfaces at different conditions
$$
r\propto t^{\alpha}, \quad 1/10 \le \alpha \le 1/6.
$$

Still this rate is slower than that found in the spreading experiments over calm waters in an axisymmetric case with the exponent $\alpha=1/4$~\cite{Hoult1972}. In a one-dimensional case, $n=1$, the situation is similar. The self-similar solution implies $\alpha=1/5$, while the viscous spreading rate has an exponent $\alpha=3/8$.

Such a trend should be expected, as the liquid substrate provides a very effective lubrication layer for the spreading liquid. Therefore, the model we utilize is not expected to accurately describe spreading over a steady interface or in the limit of very small amplitudes $B_0\ll 1$. 

\begin{figure}[ht!]
\begin{center}
\includegraphics[trim=0.3cm 1.cm 1cm -0.5cm,width=0.9\columnwidth]{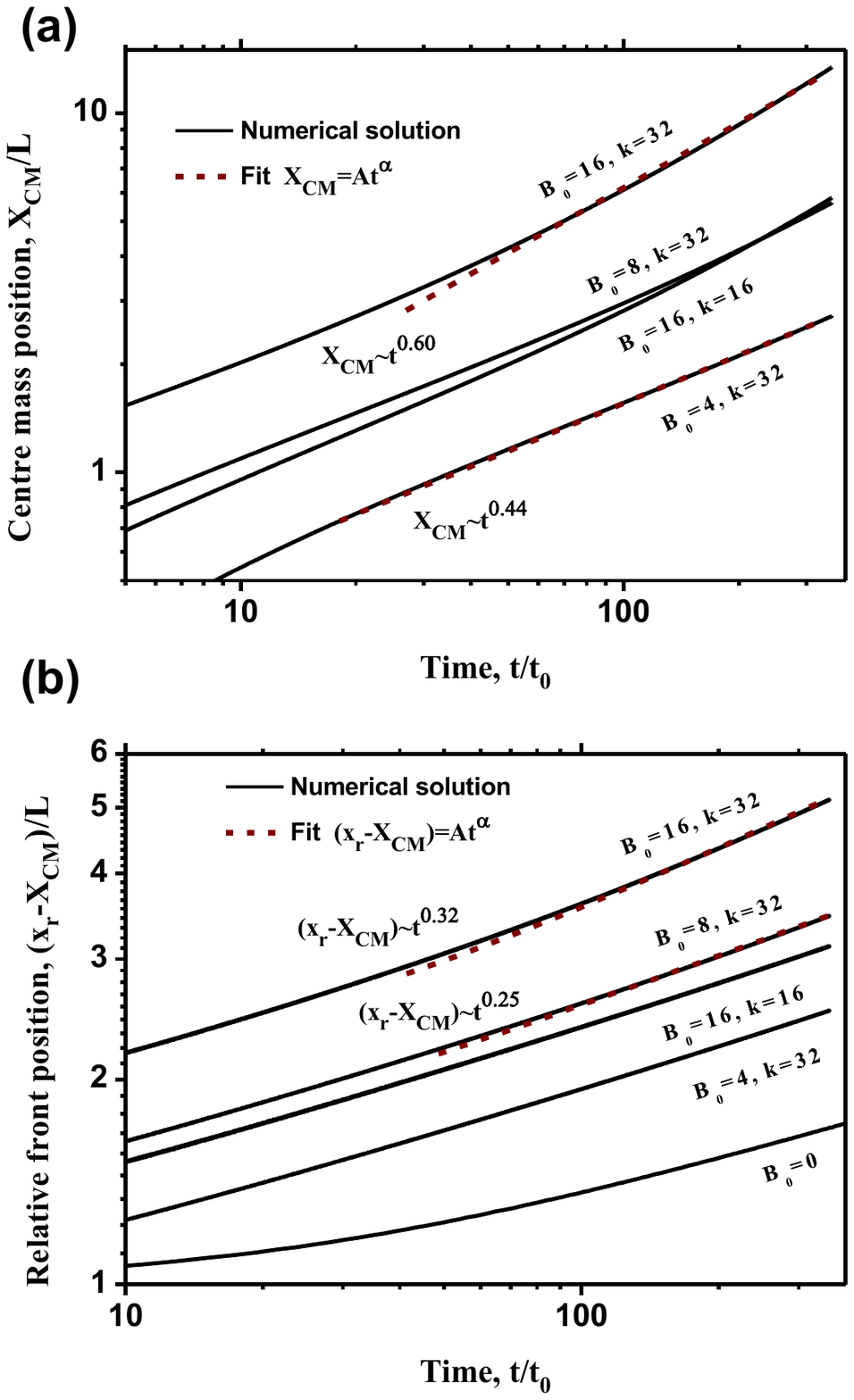}
\end{center}
\caption{Evolution of the oil slick over a wavy interface at different perturbation amplitudes $B_0$ and wave numbers $k$. (a) the position of the centre of mass $X_{CM}(t)$ of the oil slick and (b) the position of the moving front $x_r(t)-X_{CM}(t)$ initially located at $x=1$ relative to the centre of mass. The solid lines are numerical solutions and the dashed lines are the power fits $f=f_0 t^{\alpha}$. The problem parameters have been fixed at $\alpha_g=0.02$, $Ka=0.1$ and $\varepsilon=2\cdot 10^{-4}$. The initial profile (at $t=0$) was the self-similar solution (\ref{SSS}).} 
\label{Fig2}
\end{figure}

\begin{figure}[ht!]
\begin{center}
\includegraphics[trim=0.3cm 1.cm 1cm -0.5cm,width=\columnwidth]{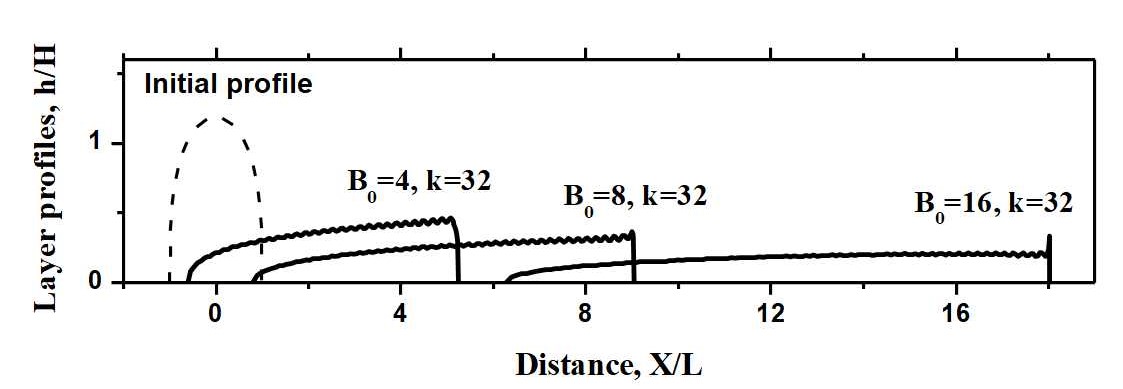}
\end{center}
\caption{Oil layer profiles $h(x,t)$ at the end of the evolution shown in Fig.\ref{Fig2}  (at $t/t_0=350$) at different perturbation amplitudes $B_0$. The initial profile was the self-similar solution (\ref{SSS}).} 
\label{Fig3}
\end{figure}

\begin{figure}[ht!]
\begin{center}
\includegraphics[trim=1.1cm 1.cm 0.cm -0.5cm,width=0.8\columnwidth]{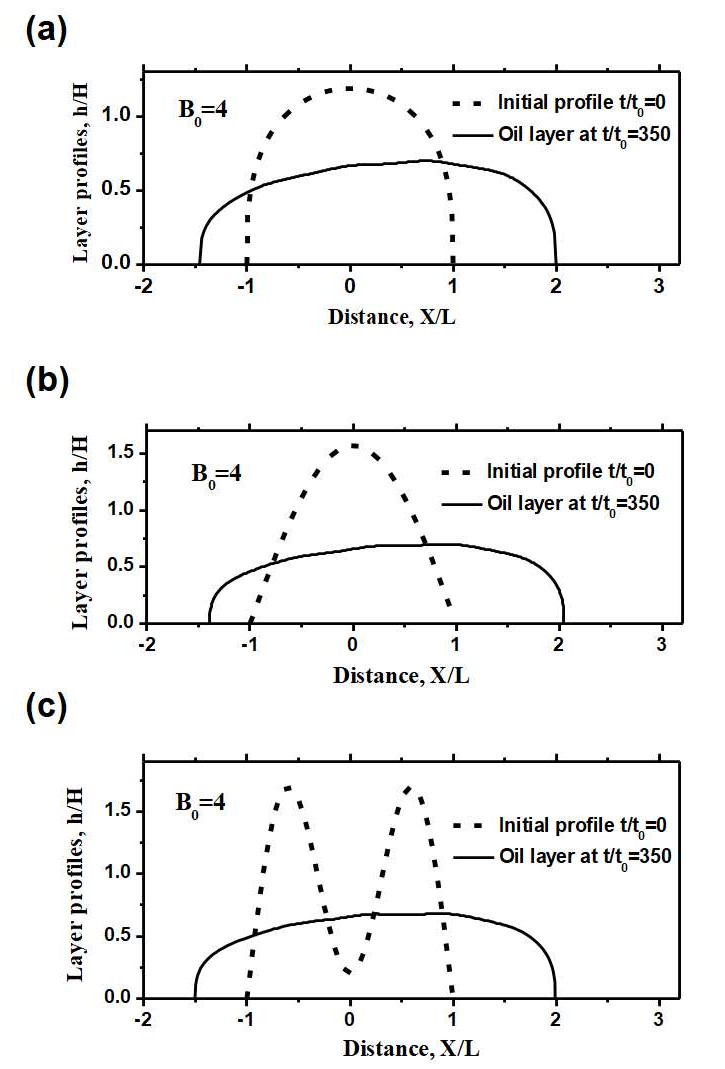}
\end{center}
\caption{Oil layer profiles $h(x,t)$ at the beginning (dashed lines) and the end (solid lines) of the evolution (at $t=0$ and $t/t_0=350$ respectively) at $B_0=4$, $k=8$ ($\omega \approx 63$) and different initial profiles. The other problem parameters are as in Fig. \ref{Fig2}. The initial profile in panel (a) corresponds to the self-similar solution (\ref{SSS}).} 
\label{Fig22}
\end{figure}

\subsection*{Spreading in the presence of travelling-wave perturbations}

Consider now the dynamics when perturbations are present. We first analyze the effect of a single harmonic perturbation in the form of a travelling (in the positive direction) wave
\begin{equation}
\label{DPWTH1}
B(x,t) =  B_0\sin(k x - \omega t), 
\end{equation}
$$
V(x,t) =  B_0 \varepsilon \omega \sin(k x - \omega t).
$$

In the simulations, initially, the oil layer domain is conveniently set to be $x\in[-1, 1]$ with the total amount of the oil $M=\int_{-1}^{1}\, u\, dx = 2$ to obtain initial layer thickness of the order of one.  

To understand the effect of the travelling-wave perturbations, we consider two sets of dimensional and non-dimensional parameters corresponding to different characteristic heights of the oil layer and possibly different stages of spreading. 

In the first set of parameters, which is mostly characteristic for the initial phases of the oil spill dynamics, the oil layer thickness is taken at $H=10\,\mbox{mm}$. The horizontal length scale and the characteristic velocity in both cases are taken at $L=50\,\mbox{m}$ and $U=1\,\mbox{m}/\mbox{s}$ respectively, so that $t_0=50\,\mbox{s}$. 

If we take $\rho=8.7\cdot 10^{2}\,\mbox{kg}/\mbox{m}^3$ and $\mu=8.7\cdot 10^{-3}\,\mbox{Pa}\cdot\mbox{s}$ corresponding to light oil, the Reynolds number is $Re=0.2$, $Ka=0.1$ and $\alpha_g=0.02$. That is $Re\ll 1$ and the assumptions are fulfilled.

In the second set of parameters, which is more common for later stages of spreading, the characteristic height is taken at $H=1\,\mbox{mm}$ giving 
$Re=0.002$, $Ka=0.01$ and $\alpha_g=2\cdot 10^{-5}$. In general, the second set of parameters can be categorised as having relatively low contribution from the diffusion, as parameter $\alpha_g$ is much smaller, emphasising the role of the convective terms. 

In what follows, the problem (\ref{LE2TH1D}) was solved numerically using a moving-mesh method~\cite{Baines2015}. The spatial resolution was set in the range from 100 to 200 mesh-points for low $k$-number ($k\le 16$) simulations and from 200 to 4000 mesh-points for high $k$-number ($k>16$) simulations, while the time step was adjusted to achieve numerical stability. For the sake of comparison, we set the initial profile by default according to the self-similar solution (\ref{SSS}), unless otherwise stated, but have taken several test runs, where the initial profile varied.

\begin{figure}[ht!]
\begin{center}
\includegraphics[trim=0.3cm 2.cm 1cm -0.5cm,width=0.9\columnwidth]{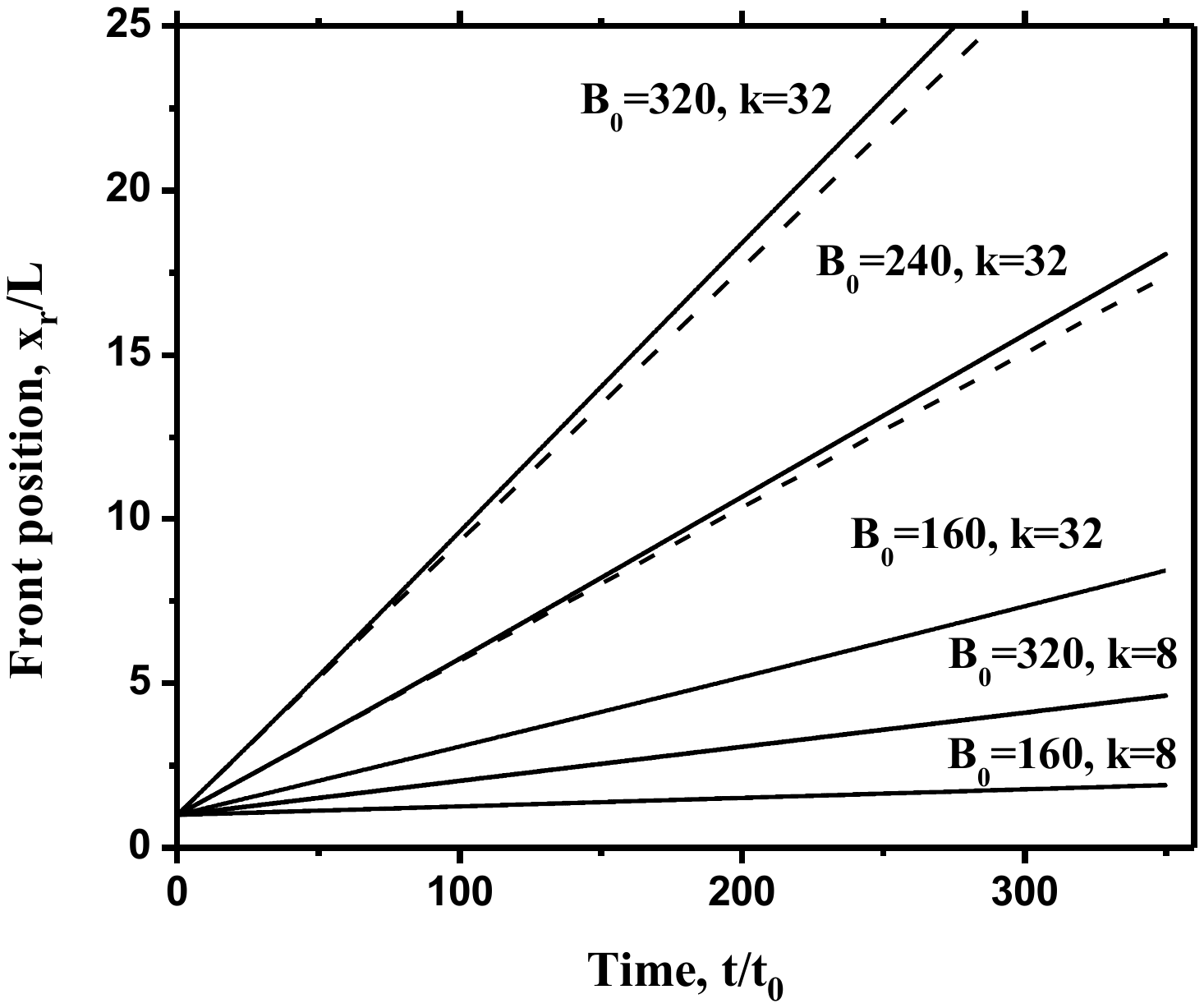}
\end{center}
\caption{Evolution of the moving front $x_r(t)$ initially allocated at $x=1$ over a wavy interface at different perturbation amplitudes $B_0$ and the wave numbers $k$. The other problem parameters were fixed at $\alpha_g=2\cdot 10^{-5}$, $Ka=0.01$ and $\varepsilon=2\cdot 10^{-5}$. The initial profile was the self-similar solution (\ref{SSS}). The dashed lines are the evolution curves with velocities according to (\ref{StokesD}).} 
\label{Fig4}
\end{figure}

\begin{figure}[ht!]
\begin{center}
\includegraphics[trim=0.3cm 1.6cm 0cm -0.5cm,width=0.8\columnwidth]{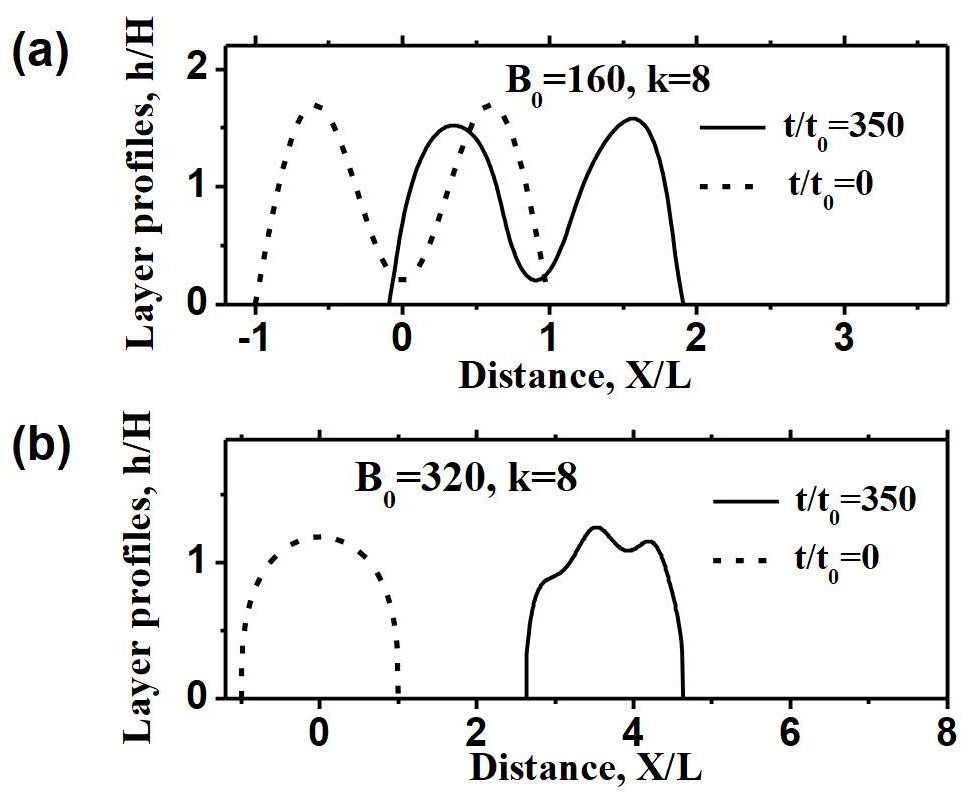}
\end{center}
\caption{Evolution of the oil layer profiles $h(x,t)$ in advection dominated regime illustrated in Fig.~\ref{Fig4} at low wave numbers $k$ at different perturbation amplitudes $B_0$ and different initial profiles at $\alpha_g=2\cdot 10^{-5}$, $Ka=0.01$ and $\varepsilon=2\cdot 10^{-5}$. The initial profile in (b) was the self-similar solutions (\ref{SSS}).} 
\label{Fig231}
\end{figure}

\subsubsection*{Diffusion dominated regime}

Using the first set of parameters, the evolution of the oil slick has been studied parametrically by varying the amplitude of the perturbations $B_0$, the wave number $k$ and in some cases the initial conditions. The results are illustrated in Figs. \ref{Fig2}-\ref{Fig22}. 

As the first example, we consider perturbations at $k=32$ and, according to (\ref{DispR}), $\omega\approx 127$. The evolution of the oil slick shape is shown in Fig. \ref{Fig3}. As one can observe, the spreading process can be described as simultaneous drift and diffusion. While the total length of the slick is becoming larger, both boundaries move in the direction of the wave group velocity. One may note then that the observed trends in the numerical simulations are qualitatively in accord with direct observations~\cite{Ermakov2018}.

In the analysis, the evolution of the oil slick profile is split into two parts, the motion of the centre of mass $X_{CM}=\int_{x_l}^{x_r}\, h x\, dx \, M^{-1}$ and the motion of the boundaries, in particular $x_r(t)$ with respect to the centre of mass $x_r-X_{CM}$.

To make a comparison with the experiments conducted on the surface of calm water~\cite{Hoult1972}, we have provided a power law fit to the data in the form $x_r(t) - X_{CM}=A_p t^{\alpha}$, though we note that strictly speaking no self-similar behaviour is expected, Fig. \ref{Fig2} (b).  As one can see, while the wave action produces almost the same power law exponent $\alpha\approx 0.32$ as that $\alpha\approx 0.38$ found in the experiments on flat water surfaces, it can nevertheless substantially facilitate the process of spreading. The effect is roughly proportional to $(B_0 k)^{1/2}$, so that at relatively small perturbation amplitudes (starting from just a few oil layer widths), the spreading rate can be several times larger than that in the absence of the wave motion.

In terms of dimensional values, if the amplitude of the water waves is just about $4\,\mbox{cm}$, the initial domain of $100\,\mbox{m}$ spreads out to $300\,\mbox{m}$ in about $4$ hours, and if the amplitude increases to $16\,\mbox{cm}$, which is still pretty low, the domain spreads out during the same time to $600\,\mbox{m}$. 

Consider now the motion of the centre of mass. This would be informative to make a comparison with the trends expected due to the Stokes drift velocity~\cite{Higgins1953, Falkovich1958}. In the non-dimensional form, the drift velocity can be presented as
\begin{equation}
\label{StokesD}
u_s=\frac{1}{2}\omega k B_0^2 \varepsilon^2
\end{equation}
implying that $u_s\propto B_0^2 k^{3/2}$~\cite{Falkovich1958}. That is, smaller wavelength or larger frequency and larger wave amplitude should facilitate the drift. 

Relationship (\ref{StokesD}) can be obtained from a solution to the evolution equation
$$
\frac{dx}{dt} = u_0 \sin(kx-\omega t),
$$
which in our case corresponds to the evolution of the moving boundary with velocity (\ref{BVR}) in the limit of low diffusion, $\alpha_g=0$, at $u_0 = B_0 \omega \varepsilon$ from (\ref{DPWTH1}). A solution to this equation with initial position $x_0$ can be obtained by iterations assuming small parameter $\frac{u_0k}{\omega}\ll 1$. That is 
$$
\begin{array}{l}
\displaystyle
x(t)\approx x_0 + \frac{u_0}{\omega}\cos(kx_0 - \omega t) - 
\\
\displaystyle \frac{1}{4}\frac{ku_0^2}{\omega^2} \sin(2kx_0 - 2\omega t) +  \frac{1}{2}\frac{ku_0^2}{\omega} t, 
\end{array}
$$
where the last,  secular term with linear dependence on time appears in the second approximation and provides the estimate for the drift velocity in (\ref{StokesD}).

As one can see from Fig. \ref{Fig2} (a), the time dependencies of the centre of mass are non-linear, which is likely due to variations of the oil slick shape and dimensions due to the presence of diffusion. One can readily observe the trends, which are qualitatively consistent with (\ref{StokesD}). First of all, the velocity of the centre of mass increases with increasing the amplitude of the perturbations $B_0$, though the effect scales linearly with $B_0$ rather than quadratically. 

\begin{figure}[ht!]
\begin{center}
\includegraphics[trim=0.3cm 1.6cm 0cm -0.5cm,width=0.8\columnwidth]{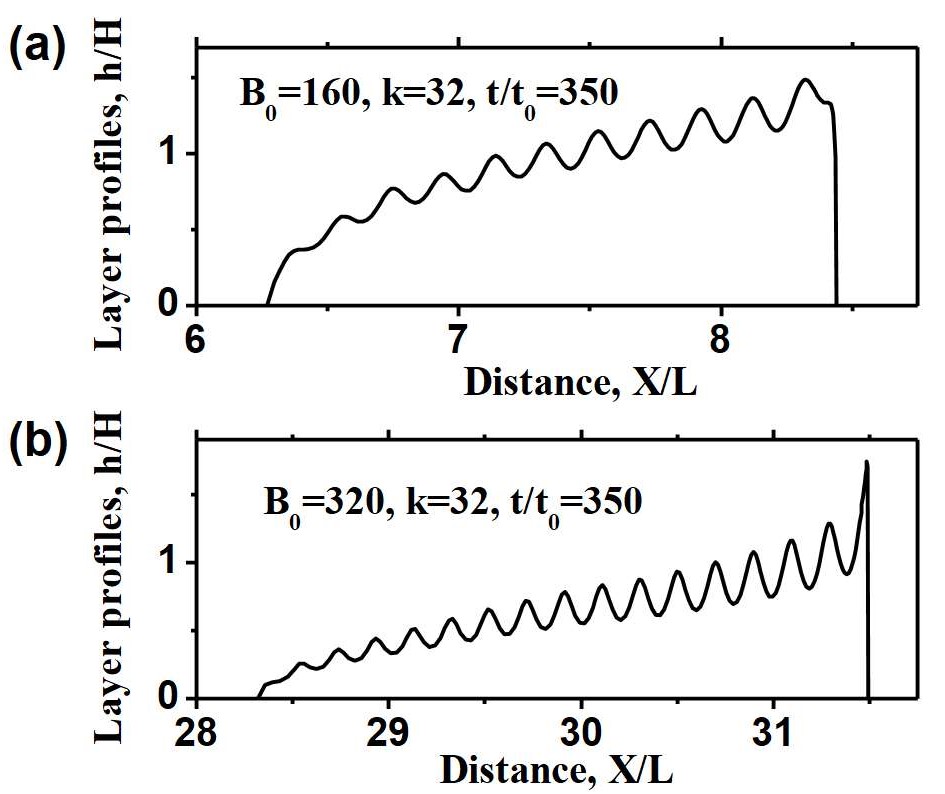}
\end{center}
\caption{Evolution of the oil layer profiles $h(x,t)$ in advection dominated regime illustrated in Fig.~\ref{Fig4} at high wave numbers $k$ at different perturbation amplitudes $B_0$ at $\alpha_g=2\cdot 10^{-5}$, $Ka=0.01$ and $\varepsilon=2\cdot 10^{-5}$. The initial profiles were the self-similar solutions (\ref{SSS}).} 
\label{Fig232}
\end{figure}

A similar effect is observed with increasing the wave number $k$, Fig. \ref{Fig2} (a). That is the short wave-length perturbations produce stronger effect on the drift of the initial profile, though again, the position of the centre of mass scales much weaker than $k^{3/2}$ as it would be expected from (\ref{StokesD}). 

Consider now how the choice of the initial profile (keeping the total amount constant, of course) may affect the slick dynamics in this diffusion dominated regime. The initial profiles have been varied at fixed values of the water-wave amplitude $B_0$ and the wave length, Fig. \ref{Fig22} . As one can observe, the effect is minimal. If we look at the profiles at the end of the evolution shown in Fig. \ref{Fig22}, one can observe that despite quite different initial conditions, the profiles quickly evolve to a universal profile resembling, though not entirely, the self-similar solution. 

\subsubsection*{Advection dominated regime}

Consider now the results obtained using the second set of parameters, when the diffusion rate is expected to be much smaller, so that convective motion should prevail. To bring the second case to an equivalent one for comparison, we roughly set the amplitudes of the perturbations to the same dimensional values. The simulation results are demonstrated in Figs. \ref{Fig4}--\ref{Fig232}. 

First of all, one can observe that the moving front is propagating linearly in time, Fig. \ref{Fig4}. Consider the evolution of the profiles at low $k$-numbers, at $k=8$, Fig. \ref{Fig231} (a) and (b). As one can see, the oil slick is drifting away from its initial position having almost initial profile, that is moving as a whole. 

 \begin{figure}[ht!]
 \begin{center}
\includegraphics[trim=0.3cm 0.3cm 1cm -1.5cm,width=0.8\columnwidth]{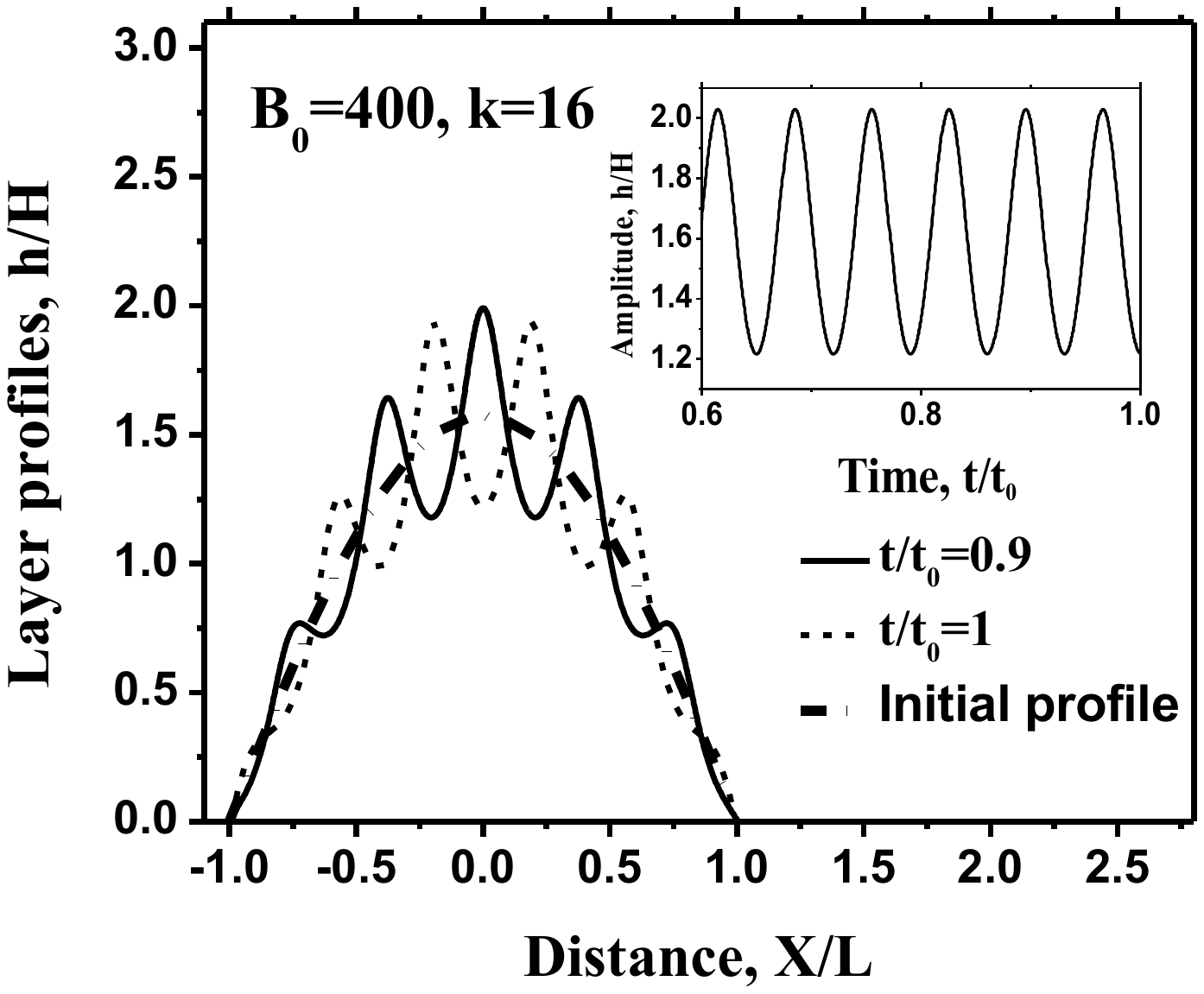}
 \end{center}
 \caption{Oil layer profile $h(x,t)$ under the action of standing water-waves at different moments of time and at $k=16$ and $B_0=400$. The inset shows the amplitude of $h(x,t)$ as a function of time at $x=0$. The model parameters $\alpha_g=2\cdot 10^{-5}$, $Ka=0.01$ and $\varepsilon=2\cdot 10^{-5}$.  The initial profile (dashed-dot line) $h(x,0)=\frac{\pi}{2}\cos(\pi x/2)$.} 
\label{Fig53T}
\end{figure}

The profile undergoes some changes at high $k$-number perturbations resembling the profiles observed in the first case, but the width is not changing much, Fig. \ref{Fig232} (a) and (b). So that again, the slick is moving as a whole.  

The rate of the front motion, which is noticeably linear with time, is proportional to the square of the wave amplitude $B_0^2$ and to $k^{3/2}$ as is expected from (\ref{StokesD}). Moreover, the correspondence is quantitative as one can see from the front rate calculated by (\ref{StokesD}) and shown for comparison in Fig. \ref{Fig4}.

One should note that the dynamics in this case is practically independent of parameter $\alpha_g$, which can be set to zero.

One can conclude then, that under the action of travelling water-wave perturbations, 'thick' oil slicks tend to drift away and spread over, while 'thin' oil layers tend to drift away preserving the initial width and possibly even the initial shape. That is the effect of the travelling perturbations in the advection regime is roughly equivalent to the Stokes drift motion of a particle when the oil slick is moving as a whole with the Stokes drift velocity.

The effects in both regimes increase with increasing the amplitude of perturbations and their wave number $k$. Apparently, the dynamic effects should be sensitive to the spectral properties of the perturbations such as the phase shifts, so that this could be interesting to see how the water-wave perturbations affect the slick dynamics in two-dimensional motion.  

 \begin{figure}[ht!]
 \begin{center}
\includegraphics[trim=0.3cm 0.3cm 1cm -1.5cm,width=0.7\columnwidth]{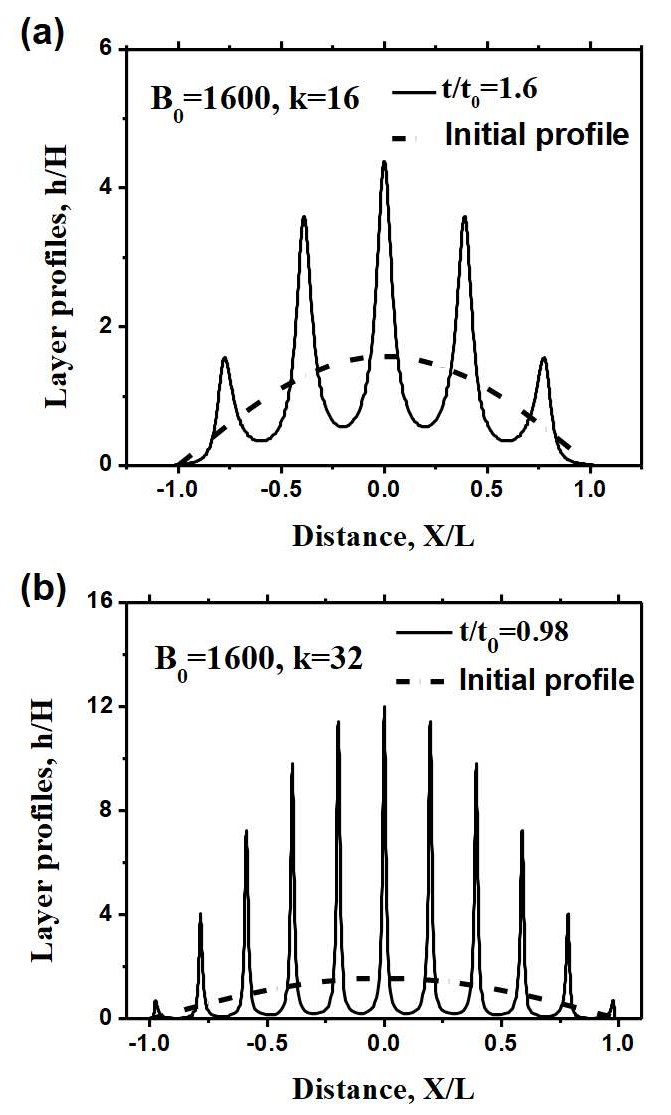}
 \end{center}
 \caption{Oil layer profiles $h(x,t)$ under the action of standing water-waves at different wave numbers $k$. The model parameters $\alpha_g=2\cdot 10^{-5}$, $Ka=0.01$ and $\varepsilon=2\cdot 10^{-5}$. (a)  $B_0=1600$, $k=16$, $\omega\approx 89$ (b) $B_0=1600$, $k=32$, $\omega\approx 127$. The initial profile (dashed-dot line) was $h(x,0)=\frac{\pi}{2}\cos(\pi x/2)$.} 
\label{Fig53}
\end{figure}

\subsection*{Spreading in the presence of standing-wave perturbations}

As we have established so far, the travelling-wave disturbances, in a one-dimensional case, facilitate the drift and rather homogeneous spreading of the oil layer. 

Another type of behaviour is observed when the water-wave motion is a standing wave 
\begin{equation}
\label{DPWTH2}
\begin{aligned}
B(x,t) =  B_0\{\sin(k x - \omega t) + \sin(k x + \omega t)\}=\\2B_0\sin(k x)\cos(\omega t), 
\end{aligned}
\end{equation}
$$
V(x,t) =  2 B_0 \varepsilon \omega \sin(k x)\cos(\omega t).
$$

 \begin{figure}[ht!]
 \begin{center}
\includegraphics[trim=0.3cm 0.3cm 1cm -1.5cm,width=0.7\columnwidth]{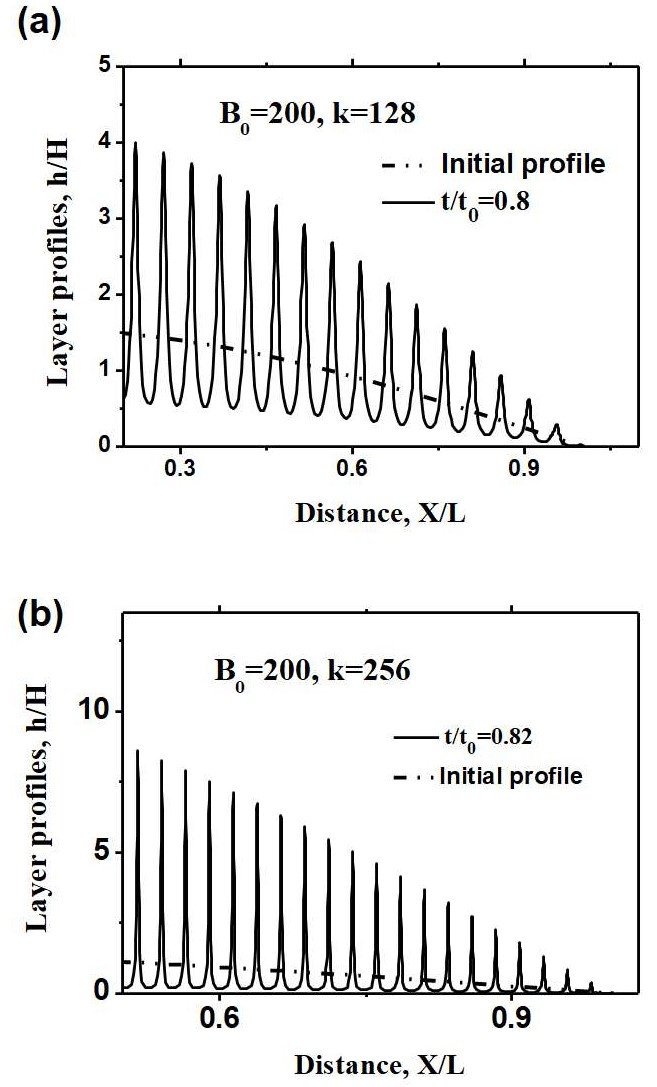}
 \end{center}
 \caption{Oil layer profiles $h(x,t)$ under the action of standing water-waves at different wave numbers $k$. The model parameters $\alpha_g=2\cdot 10^{-5}$, $Ka=0.01$ and $\varepsilon=2\cdot 10^{-5}$. (a)  $B_0=200$, $k=128$, $\omega\approx 253$ (b)  $B_0=200$, $k=256$, $\omega\approx 358$. The initial profile (dashed-dot line) was $h(x,0)=\frac{\pi}{2}\cos(\pi x/2)$.} 
\label{Fig531}
\end{figure}

To illustrate the effect, it is informative to consider the second set of parameters, when the advection processes dominate. The results of simulations with cosinusoidal, $h(x,0)=\frac{\pi}{2}\cos(\pi x/2)$, initial profiles are shown in  Figs. \ref{Fig53T}--\ref{Fig531}. 
 
 \begin{figure}[ht!]
 \begin{center}
\includegraphics[trim=0.3cm 0.3cm 1cm -1.5cm,width=0.7\columnwidth]{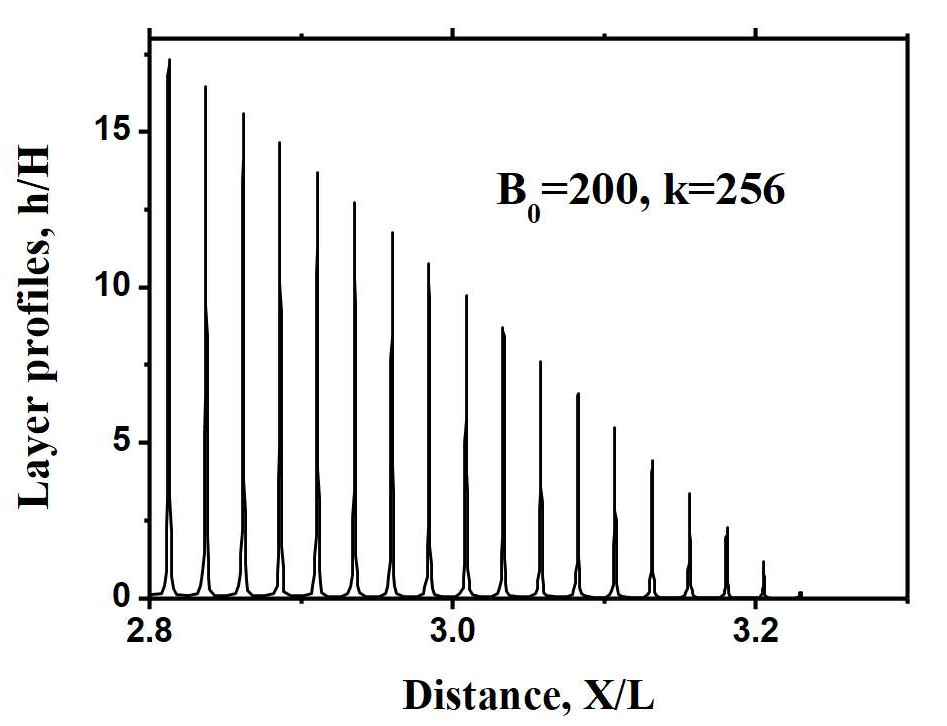}
 \end{center}
 \caption{Oil layer profiles $h(x,t)$ under the combined action of the water waves (\ref{DPWTHC}) at $t/t_0=2.5$, $k=256$ and $B_0=200$. The model parameters $\alpha_g=2\cdot 10^{-5}$, $Ka=0.01$ and $\varepsilon=2\cdot 10^{-5}$. The initial profile was $h(x,0)=\frac{\pi}{2}\cos(\pi x/2)$.} 
\label{Fig63}
\end{figure}

The dynamics of the oil layer in this case is in some contrast to the response to the travelling wave perturbations. As one would anticipate, no drift motion was observed, but the depletion of the initial profile at the divergence points of the velocity profile where $V=0$ and $\frac{\partial V}{\partial x}>0$ (at $t=0$, the locations are $x_{min}=2\pi n/k$, $n=0,\pm 1, \pm2, ...$) accompanied by simultaneous enhancement at the convergence points where $V=0$ and $\frac{\partial V}{\partial x}<0$ (at $t=0$, the locations are $x_{max}=\pi (2n+1)/k$, $n=0,\pm 1, \pm2, ...$). 

The positions of the convergence and divergence points interchange depending on the moment of time, if either $\cos\omega t>0$ or $\cos\omega t<0$, see Fig. \ref{Fig53T}. The motion of any point of the profile in this parameter range in the domain is simply harmonic with the angular frequency of the external perturbations $\omega$, see the inset in Fig. \ref{Fig53T}. 

Variations of the water-wave parameters demonstrate that as the amplitude of the water waves increases and/or the wavelength decreases, the depletions approach almost complete rupture of the oil layer, when the thickness of the oil layer becomes extremely low Figs. \ref{Fig53} --\ref{Fig531}. 

Therefore, this scenario can potentially serve to provide a mechanism of oil layer rupture practically at the initial phases of oil spreading well before any surface tension effects may become dominant and important. The characteristic time of the depletion development is on the scale of the wave motion $\omega^{-1}$.

As one can observe, the depletion effect is mostly pronounced at the boundary of the oil layer, which should be potentially a facilitating factor of the emulsification processes. At high $k$-numbers, one requires only relatively modest water wave amplitudes, on the scale of ten centimeters in dimensional terms, to achieve deep depletions, Fig. \ref{Fig531}.   

In real conditions, the surface wave disturbances consist of a spectrum of harmonics, so this would be interesting to see the layer response to the combined action. 

To mimic such conditions, one can apply a slightly unbalanced combination of two harmonics 
\begin{equation}
\label{DPWTHC}
\begin{aligned}
B(x,t) =  B_0\left\{\sin(k x - \omega t) + \frac{1}{2}\sin(k x + \omega t)\right\}, 
\\ V(x,t) =   B_0 \varepsilon \omega \left\{  \sin(k x - \omega t) + \frac{1}{2}  \sin(k x + \omega t)\right\}.
\end{aligned}
\end{equation}

The result is demonstrated in Fig. \ref{Fig63}, where one can observe that effectively the layer response consists of depletions advected downstream by the wave group velocity. So that the effect is indeed a combination of two processes, advection and depletion.

\section{Conclusions}

The dynamics of oil slicks, under the influence of surface water waves, have been quantitatively analyzed using a thin film model. This analysis identifies two characteristic regimes of oil slick spreading, contingent on the wave type: traveling or standing wave perturbations. We demonstrate that surface wave disturbances can fracture continuous oil layers during intermediate phases of the spreading process.

Specifically, traveling waves induce a drift motion, leading to a relatively uniform spreading of the oil layer. The impact is modulated by the non-dimensional model parameter $\alpha_g = \frac{\rho g_0 \varepsilon H^2}{\mu U}$, which quantifies the contribution of diffusion terms, and also by the amplitude of perturbations $B_0$ and the wave number $k$.

For very low $\alpha_g$ values, advection effects predominate, causing the oil layer to drift as a whole in the direction of the wave group velocity. This drift resembles the motion of a single particle with Stokes drift velocity, suggesting that 'thin' oil layers under modest traveling water wave perturbations will drift away, maintaining their width and initial profile.

At higher $\alpha_g$ values, where diffusion processes are more significant, the drift motion diverges from the Stokes drift paradigm, and the oil slick thickness profile approaches a universal shape akin to self-similar solutions (\ref{SSS}).

In both scenarios, the rate of spreading accelerates with the amplitude $B_0$ and the wave number $k$, amplifying the impact of short wavelength perturbations.

Conversely, standing wave perturbations produce a more complex effect. At low $\alpha_g$ values, almost immediately with the wave motion's onset, the oil layer forms pronounced peaks and troughs. Their amplitude and depth escalate with $B_0$ and $k$, potentially reducing the oil layer thickness to zero. This thinning mechanism could act as a precursor to or a mechanism for layer breakup, notably before the surface tension phase of spreading, indicating a structure prone to emulsification.

The studied parameter range aligns with the initial stages of an oil spill, when the layer's thickness remains in the millimeter scale, and prior to the predominant influence of surface tension on spreading.

In real-world conditions, surface wave perturbations exhibit both traveling and standing components, leading to a fragmented oil layer that drifts and morphs according to wave motion. From a practical standpoint, investigating two-dimensional motion with more complex (non-linear) surface wave inputs would be of interest.

\end{document}